\documentclass[12pt]{iopart}

\usepackage{iopams}                  
\usepackage{graphics}
\usepackage{graphicx}                                                                     
                                                                
\usepackage{tabularx}                                                                     
\usepackage{lscape}                                                                       
\usepackage{}                                                                             
                                                                 
\input colordvi
\usepackage[usenames]{color}
\usepackage[normalem]{ulem}

\begin{document}

\title[Surface stress of 
  Ni adlayers on W(110)]{Surface stress of 
  Ni adlayers on W(110): the critical role of the surface atomic structure }

\author{N.~Stoji\'{c}$^{1,2}$ and  N. Binggeli$^{1,2}$}

\address{$^{1}$The Abdus Salam International Centre for Theoretical Physics,
Strada Costiera 11, 34151 Trieste Italy}
\address{$^{2}$IOM-CNR Democritos, Trieste I-34151, Italy}

\ead{nstojic@ictp.it}

\date{\today}

\begin{abstract}
Puzzling trends in surface stress were reported experimentally for 
Ni/W(110) as a function of Ni coverage. In order to explain this 
behavior, we have performed a density-functional-theory study of the 
surface stress and atomic structure of the pseudomorphic and of 
several different possible $1\times7$ configurations for this system. For 
the $1\times7$ phase, we 
predict a different, more regular atomic structure than previously
proposed based on surface x-ray diffraction. At the same time, we 
reproduce the unexpected experimental 
change of surface stress between the pseudomorphic and $1\times7$ configuration along
the crystallographic surface direction which does not undergo density 
changes. We
show that the observed behavior in the surface stress is dominated by the effect of
a change
in Ni adsorption/coordination sites on the W(110) surface.

\end{abstract}


\maketitle

\section{Introduction}

The importance of surface stress has been demonstrated  for many 
 surface processes, such as nanopatterning, surface reconstruction, 
interfacial mixing, and  
segregation \cite{Hai01,Ibachs,NarVan92,NieBesSte95,CamSie94,MenStoLoc11,NeeGodMan91,SanTiaKir09,BlaJen10}. 
In spite of its importance, however, a general understanding of the
key  factors governing the dependence of  surface stress  on the
atomic structure of the surface is still  lacking. This  applies, in
particular,  to the case of heteroepitaxy, and to the growth of metal 
adlayers on metal surfaces, where lattice mismatch is often assumed to 
be the prevailing factor controlling surface stress \cite{San99}.

Specifically, in the case of metal adlayers on metal surfaces, 
puzzling changes in surface stress, as a function of adlayer
coverage, have been observed
experimentally \cite{SanEndKir99,MulDahIba02,SanSchEnd98,MeySanPop03}.
Such changes are at variance with predictions based on lattice misfit
arguments and model elasticity theory \cite{SanEndKir99,MulDahIba02,SanSchEnd98,MeySanPop03}.
One interesting example is the Ni adlayer on W(110), whose surface
stress was  examined in some detail 
experimentally \cite{SanSchEnd98,MeySanPop03}.
In this system, the surface-stress changes along two orthogonal
directions of the tungsten surface, [1${\bar1}0$] and [001], were
measured as a function of Ni  coverage \cite{SanSchEnd98,MeySanPop03}.
With increasing coverage, Ni goes through a range of
phases \cite{KolBau84,SchSanEnd98}, a pseudomorphic (PS) $1\times1$
configuration,  $1\times8$ and $1\times7$ coincidence structures, and
finally a fcc(111)-like  Ni overlayer with Nishiyama-Wassermann
orientation on  W(110) \cite{KolBau84}. The epitaxial strain of the
Ni(111)  layer is decreasing with increasing coverage, from
$\sim27$~\% in the PS to $~\sim-1$~\% in the $1\times7$ structure
along W[001], while it  remains constant ($\sim4$~\%) in these two
phases along W[1${\bar1}0$].  The atomic density is increasing to 9 Ni
per 7 W  atoms along [001] in the $1\times7$ structure, whereas no
change in periodicity  relative to the PS configuration occurs in the
perpendicular direction \cite{SchEndSan98}. Interestingly, it was
measured that the stress  change between PS and $1\times 7$ along
[1${\bar1}0$] is  considerably greater than the one along [001],
where actually a change of strain occurs. It remains unclear what are
the microscopic changes in the Ni/W(110) atomic structure related to
this  surprising behaviour.

The structural properties of Ni/W(110) have been investigated
experimentally  by various 
techniques \cite{KolBau84,SchSanEnd98,KolBau85,KamSchGun88,KozLilBau90,CamRodGoo90,SchEndSan98,Bau99,RifFraShi08}.
The types of reconstructions are known from low-energy electron
diffraction (LEED) \cite{KolBau84,SchEndSan98} and scanning tunneling
microscopy (STM) \cite{SchSanEnd98}, but the actual details of the
atomic structures are not well known or have large uncertainties. A
model for the  $1\times7$ coincidence structure has been proposed based on  surface x-ray 
diffraction (SXRD) measurements \cite{MeySanPop03}. It is characterized
by distorted Ni hexagons and by a surprisingly large motion (up to
0.5~\AA) of the subsurface-layer W atoms. Such displacements of the W
atoms were suggested to be responsible for the anomalous behaviour of
the Ni/W stress \cite{MeySanPop03}. However, because of the coexistence
of various Ni phases at the deposition temperature used in
Ref.~\cite{MeySanPop03}, there is significant uncertainty on the
structural details derived from the SXRD \cite{MeySanPop03}.

Ni/W is a prototype bimetallic  system 
\cite{KolBau84,SchSanEnd98,KolBau85,KamSchGun88,KozLilBau90,CamRodGoo90,SchEndSan98,Bau99,RifFraShi08,FarBerLi90,LiBab92,BovRudPou01,BerGoo87,GreBerGoo87,MacWurKos95,KhaChe03,KhaChe03_2}.
Such systems  are particularly interesting due to  their catalytic
properties \cite{Cam90,Rod96,BerGoo87,GreBerGoo87,CamRodGoo90,MacWurKos95,KhaChe03,KhaChe03_2},
which often closely depend on  surface structural changes.  
In addition, Ni has a very small atomic radius (the
smallest of all fcc metals), so the study of its growth on
bcc surfaces, such as the W surface, is especially  helpful for  understanding 
fcc/bcc metal interfaces \cite{Merwes,KolBau84}.
The Ni/W system has been studied amply also for  its  interesting
magnetic properties \cite{FarBerLi90,LiFarBab90,LiBab92,BovRudPou01}.

In this paper, we address the  surface stress and atomic structure of
the Ni/W $1\times1$ and $1\times7$ phases by  means  of first-principles 
density-functional-theory (DFT) calculations. In particular, we
investigate the dependence  of the surface  stress on the atomic
structure, with the aim of better understanding the key factors
controlling the stress behaviour in this type of bimetallic systems.  
 Ni/W(110) is an interesting case
for this type of study as a  surprising and  unexplained stress change
(related to unidentified structural modifications) between the two 
phases has been measured experimentally. For 
the $1\times7$ phase we predict a different, more regular atomic structure than the one
proposed on the basis of the SXRD
analysis \cite{MeySanPop03}.  The surface stresses we obtain for the
theoretical lowest-energy $1\times1$ and $1\times7$ configurations
account well for the stress behaviour observed experimentally.
Moreover, the change in surface stress between the $1\times1$ and
$1\times7$ structure is shown to be dominated by the effect of  a change in the  Ni
adsorption/coordination sites on the W(110) surface. This is in
contrast to the situations where surface stress is determined mostly
by  elastic strains or charge transfer, which can be  excluded
in this system.

\section{Methods}

The DFT calculations were performed using pseudopotentials and a plane-wave
basis, as implemented in the PWscf code, a part of the Quantum
ESPRESSO  distribution \cite{GiaBarBon09}. 
The local-density approximation (LDA) in the Perdew-Zunger
parametrization \cite{PerZun81} was adopted for the exchange and
correlation functional.   To simulate surfaces with
different  adlayer configurations, we used a supercell method,  for
which  we constructed an asymmetric slab with 1~Ni layer on 5~layers
of W substrate (bottom 2~W layers were fixed) and 9~equivalent vacuum layers,
both for relaxations and subsequent stress
calculations. Laterally, we used $1\times7$ cell.
 Only for the calculation of the adsorption sites
we used 15-layer asymmetric $1\times1$ slabs, in which 12 layers were allowed to
relax.  

Vanderbilt ultra-soft pseudopotentials \cite{Van90} were generated from
the $3d^94s^1$  atomic configuration of Ni and from the  $5s^2 5p^6
5d^4 6s^2$ configuration of W.  The core-cutoff radii  for Ni were:
$r_s=2.0$~a.u. and $r_d=1.6$~a.u., and for W: $r_{s,p}=2.2$,
$r_d=2.4$~a.u.  Our kinetic energy cutoff was 35~Ry for the wave
functions and 350~Ry for the charge density. A $38\times38\times1$
k-point  Monkhorst-Pack  mesh \cite{MonPac76} centered at $\Gamma$ was
used for the  $1\times1$ surface unit cell and a grid of comparable
density $24\times6\times1$ for the $1\times7$ surface unit cell. We
employed a Gaussian level smearing of 0.01~Ry. The calculations were
performed with the theoretical W lattice constant of 3.14~\AA\, which
is only 0.6~\% different from  the experimental lattice constant of
3.16~\AA. The use of the theoretical equilibrium lattice constant
ensures that no spurious stresses are present on the
surface \cite{NieMar83}. Total energy differences were converged to
better than 0.1~meV. The forces in the $1\times7$ calculations, used
as a criterion for the structural relaxation, were converged to better 
than 2~mRy/\AA~ and the estimated  uncertainties in the positions were
0.007~\AA~ along [1${\bar1}$0] and  0.010~\AA~ along [001]. For the
clean W(110) and  Ni/W(110) PS calculations, the forces were converged
to better than 0.2~mRy/\AA~ and we estimated the uncertainties
in positions to be around 0.001~\AA~ in both directions.  

We also performed calculations for the clean
Ni(111) surface (non magnetic) which were necessary for the
  interpretation of our results. For the calculation of the Ni(111)
surface, we used our optimized lattice constant of 3.42~\AA~ and a
slab of 11 Ni layers.
From our preliminary calculations we found
the Ni PS layer on W(110) to be non magnetic; the initial
ferromagnetic Ni/W(110) configuration converged into a
non-magnetic structure. This is consistent with findings in 
\cite{GalDebAlo01} and \cite{HuaChaLeu05}. 
Hence, in the remaining part of this work all
calculations refer to non-magnetic configurations. 

Surface stresses were computed using the analytical expression 
derived by Nielsen and Martin \cite{NieMar83}, based on the 
Hellmann-Feynman theorem. The surface stress uncertainty was
estimated to be $\sim$0.3~N/m on the basis of convergence tests 
regarding the wave-function and charge-density cutoffs, the 
number of k-points, and the number of vacuum layers \cite{MenStoBin08}.
As we used asymmetric slabs (with  Ni on top and frozen W on the bottom), 
in order to obtain surface-stress values for  the Ni-terminated
surface, we subtracted the reference-surface stress of the frozen W
surface.

\section{Results and discussion}

\subsection{Pseudomorphic phase}

\begin{figure}[t]
\begin{center}
\includegraphics[width=6.5cm]{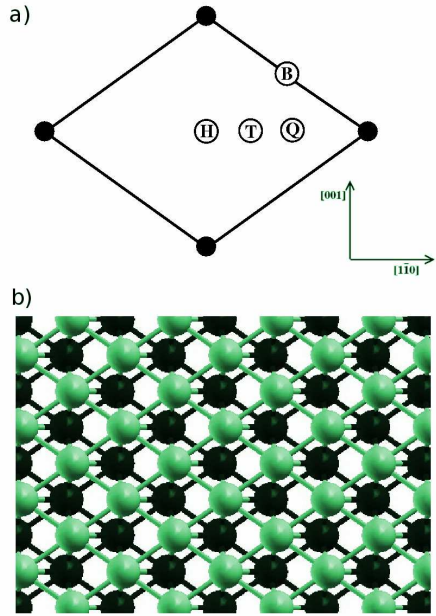}
\caption{ Possible adsorption sites for Ni/W(110) (a). T denotes
  3-fold, H  hollow, B bridge site and Q denotes the site at one
  quarter of the W-W distance along [1${\bar1}$0]. Black circles denote underlying
  W atoms, while an open circle stands for Ni atom. Calculated pseudomorphic
  structure of Ni/W(110) (b). W atoms are shown in
  black and Ni atoms in green (grey). }
\label{fig:PS_struct}
\end{center}
\end{figure}

The pseudomorphic Ni monolayer on W(110) may be viewed as a
fcc-Ni(111) layer, oriented with the in-plane Ni[1$\bar{1}$0] axis
parallel to the W[001] in-plane direction, and laterally stretched
along the W[001] and W[1$\bar{1}$0] orthogonal directions. The Ni atoms can sit
at or near any of the possible high-symmetry adsorption sites of W(110):
``hollow'' (H) in the central-symmetric position, in a perfect
continuation of the W bulk lattice,
``3-fold'' (T) for which the Ni atoms are shifted
along [1${\bar1}0$] to increase their coordination to the 3 nearest W
neighbors, or ``bridge'' (B) where the Ni atoms are located midway between two W atoms along
[$1{\bar1}1$]. The sites H, T and B are shown in
Fig.~\ref{fig:PS_struct}a). A recent  LEED I-V study \cite{note_Jacopo} 
found that Ni atoms in the PS structure prefer to sit somewhere off
the H site with a displacement toward the T-site. This is in apparent
contrast to the case of the PS Fe on W(110), where the Fe atom has
been reported to sit at the H site \cite{TobYnzPal97}.
 
Table~\ref{table:PS_displacements} collects our calculated data on 
energy differences and relaxed atomic structures  corresponding  to
initial configurations with the Ni atom at the H, T and B
sites. The relaxed T' structure, obtained starting from the Ni at the
T site, has the lowest-energy, albeit with a very small energy 
difference with respect to the H site. This small energy difference
and rather large displacement from the H site create a flat energy profile
over a significant  distance along [1${\bar1}0$] (H-T').  This energy
profile is likely to result in soft modes, which can persist also at
higher temperatures.  We note that the same relaxed
T' structure is obtained starting from a configuration where the Ni is
at the Q site, which is at 1/4 of the W-W distance along
[1${\bar1}$0] [Fig.~\ref{fig:PS_struct} (a)] and from a lower-symmetry configuration
where  the Ni atom is located  at an intermediate position
between the H and B site.  The B site is clearly very 
unfavorable ($\sim 0.36$~eV per Ni surface atom higher than T'). Figure~\ref{fig:PS_struct} b)
shows the lowest-energy pseudomorphic structure (T').  The
Ni is close to the H site, but displaced along [1${\bar1}$0]
at slightly less than 1/2 of the H-T distance.  
All results in  Table~\ref{table:PS_displacements} are for
the 15-layer asymmetric slabs (the bottom 3 layers are fixed to the bulk positions,
while the remaining 12 layers are allowed to relax). We note that
the $\Delta x$
displacements in  Table~\ref{table:PS_displacements} are given with
respect to the  bulk coordinates.

Our results are thus
consistent with the slight displacement of the H site observed by LEED
I-V.  The T' site  for Ni/W(110) may come as a surprise if one
assumes a strong influence of the substrate in determining the
adsorption site.  Actually, for unreconstructed one-monolayer
adsorbate surfaces it is often simply assumed that the adsorption will
occur as a continuation of bulk. On the other hand, one
can reason that Ni(111) has a hexagonal in-plane structure with a 3-fold
coordination, which might influence a shift away from the hollow site.

\renewcommand{\baselinestretch}{1}
\begin{table}[th]
\caption{Energy difference between the equilibrium pseudomorphic
  configurations obtained by structural relaxations starting from the
  Ni atom at the  H, T and B adsorption sites, respectively.  The
  relaxations of  the interlayer distances ($\Delta d_{n,n+1}$)  and, for
  the T' configuration, the displacements  along [1$\bar{1}$0] ($\Delta $x), 
relative to the ideal hollow site (or the ideal bulk sites for the W)
are also reported. $\Delta E$ is in meV per Ni surface
atom, $\Delta x$  in \AA, while  $\Delta d_{n,n+1}$ is given in \% of
the bulk W-W interlayer spacing.  }
\begin{center}
\begin{tabular}{ |c|c|c|c|c|}
\br
 &\multicolumn{1}{c|}
 {${\rm H}$}&\multicolumn{2}{c|}{${\rm T'}$}&\multicolumn{1}{c|}
 {${\rm B}$}\\
\cline{1-5}
  $\Delta E$ &\multicolumn{1}{c|}
 {$3$}&\multicolumn{2}{c|}{$0$}&\multicolumn{1}{c|}
 {$359$}\\
\cline{1-5}
& $\Delta d_{n,n+1}$ &$\Delta x$ &$\Delta d_{n,n+1}$  &$\Delta d_{n,n+1}$   \\
\hline  
 $Ni$       & $-18.9$    & $0.26$ &$-18.6$ & $-11.4$  \\
 $W-1^{st}$  & $0.4$ & $0.00$  & $0.3$  & $0.0$ \\ 
 $W-2^{nd}$  & $0.1$ & $0.01$  & $-0.1$  & $0.0$\\
 $W-3^{rd}$  & $0.1$  & $0.01$  & $0.0$  & $0.0$\\
\br
\end{tabular}
\label{table:PS_displacements}
\end{center}
\end{table}

\subsection{$1\times7$ coincidence phase}

\begin{figure}[t]
\begin{center}
\includegraphics[width=12cm]{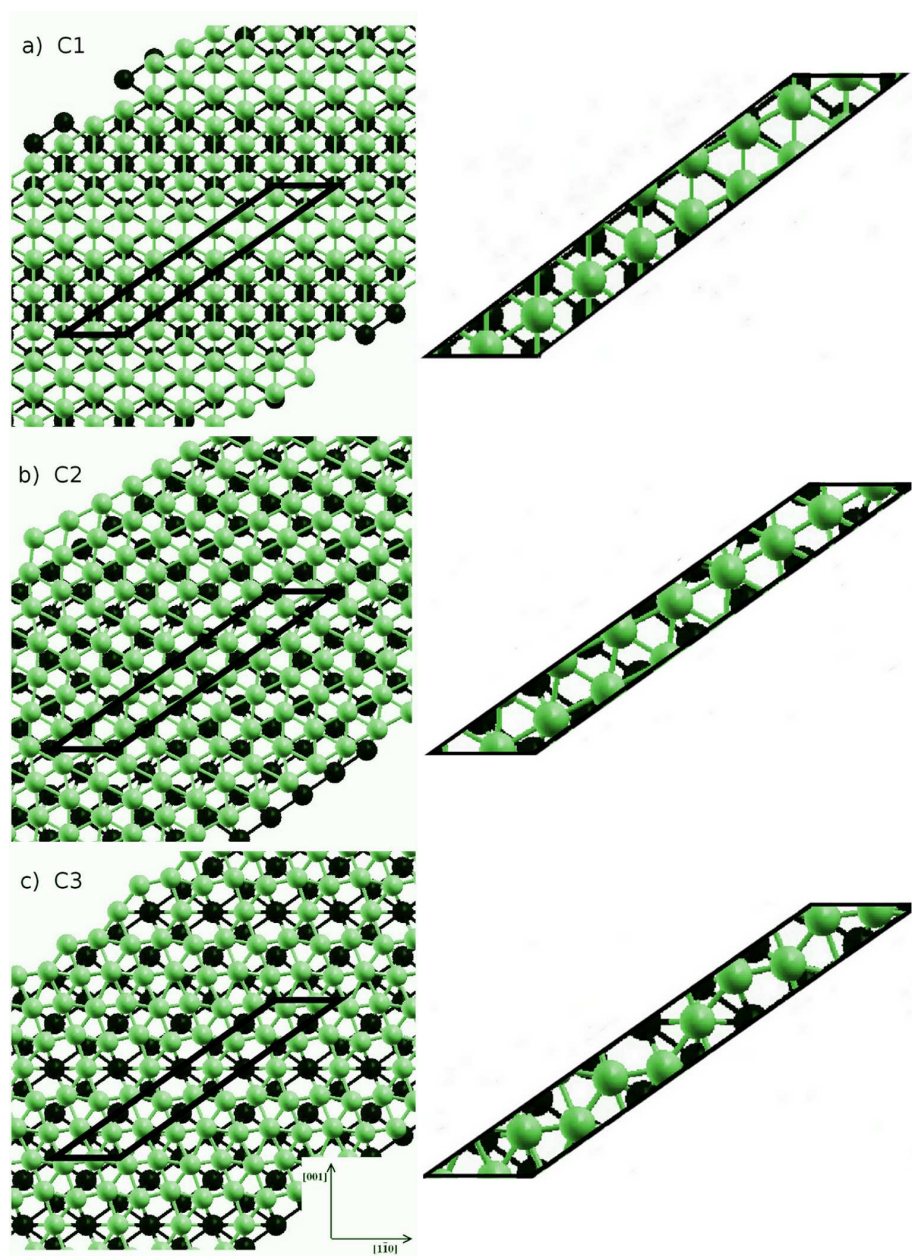}
\caption{ Relaxed 1x7 structures of Ni/W(110). Atomic colors and
  directions are like in Fig.~\ref{fig:PS_struct}. The surface unit
  cells are also indicated (thick black solid lines) 
  and shown enlarged in the right column for each structure.}

\label{fig:1x7_struct}
\end{center}
\end{figure}

The Ni/W(110) $1\times 7$ structure is characterized by higher Ni
density than the PS. As established experimentally \cite{SchEndSan98},
it has 9 Ni atoms per 7 substrate unit cells (the corresponding unit
cell is illustrated in Fig.~\ref{fig:1x7_struct}). The 
additional atoms are along the [001] direction, while along [1${\bar1}0$] 
it remains pseudomorphic. 

We obtained three different stable and meta-stable solutions depending on the initial
configuration for the structural relaxation. 
They are shown in Fig.~\ref{fig:1x7_struct}. Starting
from an ``ideal'', equidistant Ni-row arrangement along [001] and [1$\bar{1}$0], configuration C1 
[Fig.~\ref{fig:1x7_struct} (a)] has been obtained.  
The C2 structure [Fig.~\ref{fig:1x7_struct} (b)] was obtained starting from
several different initial structures, including the SXRD structure
from \cite{MeySanPop03}. The 
C3 structure [Fig.~\ref{fig:1x7_struct} (c)] resulted instead from a
relaxation starting from a structure similar to the model superstructure of
Ref.~\cite{MeySanPop03}, but slightly less disordered, keeping
only the distorted Ni hexagons. 
The C2 is the ground state, it is 
196~meV and 244~meV per $1\times1$ W surface unit lower than the C3 and
C1 configurations respectively. 

The Ni layer in the C1 structure is
very corrugated (with Ni sublayer separations, $\Delta z$, as large as
0.57~\AA), as some of the Ni atoms sit on top of the underlying
W atoms. The C2 structure [Fig.~\ref{fig:1x7_struct} (b) and atomic
coordinates in Appendix] can be described by 
oscillating chains of Ni atoms, oriented parallel to the W[001]
direction, and with their axis projection on the W surface  located
mid-way between adjacent [001] rows of W  atoms. The Ni layer
in the C2 structure is corrugated only
slightly ($\Delta z$ less than
0.15~\AA) and is very similar to the bulk Ni(111) layer - same average
nearest-neighbor (NN) distance of 2.51~\AA, with a root-mean-square
(rms) deviation of 0.12~\AA, and an angular rms deviation of
4.9$^\circ$ with respect to the average NN angle of 60$^\circ$.
The C3 structure, instead, as shown in
Fig.~\ref{fig:1x7_struct} (c), has a 
more complicated
structure, with distinct distorted Ni hexagons.

Based on SXRD measurements, Meyerheim et al. \cite{MeySanPop03}  proposed 
a rather disordered structure, characterized by distorted Ni
hexagons and Ni atoms located mostly at or close to the bridge sites
between W atoms.  We note that C2 does not resemble
it much.
We find from our calculations that the
bridge-site configuration is energetically very unfavorable.
A possible explanation for the discrepancy between our theoretical structure and
the SXRD-derived atomic structure is the coexistence of various phases with domain
sizes of few nm \cite{SchEndSan98,SchSanEnd98}  observed under the
growth conditions used in the SXRD measurement \cite{MeySanPop03}.
For such small domains, the structural relaxations at the domain boundaries are expected
to be significant \cite{MenStoBin08}. Furthermore, the coexistence of phases with similar periodicity,
i.e. $1\times8$ and $1\times7$ phases, combined with the small domain sizes, possibly makes it
difficult to distinguish the corresponding reciprocal-space features in the diffraction
measurements. The disorder, which was essential in fitting the diffraction data
in the mentioned SXRD study  \cite{MeySanPop03}, is consistent with these observations. We note that
such a disorder, resulting from room-temperature measurements, 
cannot be obtained from a coexistence of C1, C2 and C3 configurations
due to the large energy differences between the C2 and the other two structures.
However, we cannot exclude the possibility that the C2 structure could represent
a better model to fit the SXRD data, as the study in  \cite{MeySanPop03} does not seem to take into
account a C2-like structure in searching for the optimal fit.

\renewcommand{\baselinestretch}{1}
\begin{table}[th]
\caption{The first two interlayer distances for the $1\times7$
  configurations, compared to the PS and clean W(110). The layer "0"
  refers to Ni.  }
\scriptsize
\begin{center}
\begin{tabular}{ |c|c|c|c|c|c|}
\br
 & $C1$ & $C2$ & $C3$ & $PS$ & $W$ \\
\hline \hline  
 $\rm d_{01}$ (\%)  & $-6.1$ & $-10.1$ & $-9.5$ &  $-18.6$ & $$ \\
 $\rm d_{12}$ (\%)  & $1.1$ & $0.0$   & $0.5$   &  $0.3$ & $-3.6$ \\
\br
\end{tabular}
\label{table:d12}
\end{center}
\end{table}

Our C2 structure, instead, corresponds to and is very much alike  the
theoretical model prediction of Ref.~\cite{Merwes} for the
$1\times8$ structure of
Ni/Mo(110) and the experimental and theoretical $1\times8$ structure of 
Co/W(110) \cite{PraElmGet03,SpiHaf04}. In our case,  the
amplitude of the oscillations of the Ni chains is about 0.34~\AA. 
Mo and W are both bcc metals and have almost identical lattice
parameters (the difference is about 0.6~\%). Clearly, the model  predictions
for the Ni/Mo(110) are very relevant also for Ni/W(110), as our optimized structure
demonstrates. Similarly, the nearest neighbor distances for bulk Ni and
Co are very close ($\sim0.5$~\% difference), and one could expect some
similarities of their adsorbate structures on W(110). Therefore, based
on our results on Ni/W(110), one could expect that also the $1\times7$ phase of
Co/W(110) should exist for  coverages within the interval determined by the
$1\times1$ and $1\times8$ phases. 

In Table~\ref{table:d12} we compare the first two average interlayer distances
of the three $1\times7$ configurations with the PS and clean W(110)
surface \cite{note_W}. PS has
the largest contraction of the first layer, roughly two times larger
than the $1\times7$ configurations. Of those, C1's first layer is the
least contracted due to the on-top position of the Ni atoms in a part
of the surface unit cell.  
In all the $1\times7$ and $1\times1$ adsorbate structures considered, the W top-layer relaxation
of $-3.6$~\% is removed. Only  the C1 structure displays  a non-negligible
$d_{12}$ relaxation, due to a relatively far top layer.  
We note that, although  C2 and C3 have similar values for $d_{01}$,
C2's interlayer spacing of   2.00~\AA~ is more similar (with less than 1~\% difference) to the
first layer spacing of the Ni(111) surface, which has a very small
relaxation (we calculated it to be $-0.4$~\%).

\subsection{Surface stress}

\begin{figure}[t]
\begin{center}
\includegraphics[width=5.8cm,angle=270]{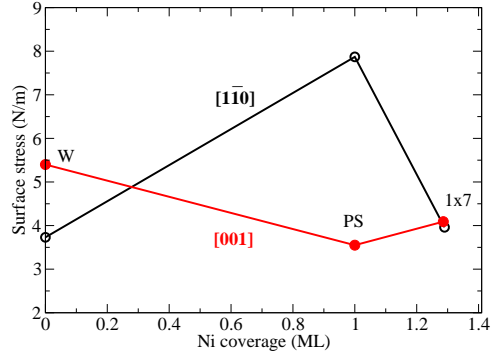}
\caption{ Evolution of the surface stress with  Ni coverage, from
  the clean  W(110) surface to the  Ni/W(110) PS 
and $1\times7$ configurations.  Open symbols, connected with a black line,
denote stress along [1${\bar1}0$] and filled symbols, connected with a
red (grey) line, along [001]. Lines are only guide to the eye.  }
\label{fig:stress}
\end{center}
\end{figure}

\renewcommand{\baselinestretch}{1}
\begin{table}[th]
\caption{Surface stress for the $1\times1$ structures characterized by different adsorption
  sites.  All stresses are in N/m. }
\scriptsize
\begin{center}
\begin{tabular}{ |c|c|c|c|c|c|}
\br
 & $H$ & $T'$ & $T$ & $Q$ & $B$ \\
\hline \hline  
 $T_{[1\bar{1}0]}$& $8.9$ & $7.9$  & $6.1$  &  $5.1$ & $5.9$ \\
 $T_{[001]}$     & $2.9$ & $3.6$  & $5.8$  &  $8.4$ & $7.5$ \\
\br
\end{tabular}
\label{table:PS}
\end{center}
\end{table}

\renewcommand{\baselinestretch}{1}
\begin{table}[th]
\caption{Surface stress for the three $1\times7$ structures. 
  All stresses are in N/m. }
\scriptsize
\begin{center}
\begin{tabular}{ |c|c|c|c|}
\br
 & $C1$ & $C2$ & $C3$  \\
\hline \hline  
 $T_{[1\bar{1}0]}$& $9.5$ & $4.0$  & $0.2$   \\
 $T_{[001]}$     & $2.7$ & $4.1$  & $3.5$   \\
\br
\end{tabular}
\label{table:1x7}
\end{center}
\end{table}

In Fig.~\ref{fig:stress} we present the calculated surface stresses,
along the W[1$\bar{1}$0] and W[001] directions, for the clean W(110)
surface and for our ground-state PS and $1\times7$ Ni/W(110)
surfaces \cite{note_shear}. They are displayed as a
function of Ni coverage. The values for the clean W(110) surface are
consistent with previous DFT values \cite{MenStoBin08,HarWooRob08}.

A first observation is that the stresses are tensile (positive) for all
the configurations, which means that the stress relaxation would cause 
contraction of interatomic distances on the surface. This is not
surprising, as tensile
stress occurs at most metal surfaces, due to a charge
redistribution upon bulk termination.  From W to PS Ni/W
the stress is increasing along  [1${\bar1}0$], whereas it is decreasing along
[001], which is somewhat surprising. Considering the lattice mismatch,
the Ni layer is under a strongly tensile (experimentally: 27~\%,
theoretically 30~\%) epitaxial strain
along [001] (and experimentally  under a 4~\% strain, theoretically
under a 6~\% strain along [1$\bar{1}$0]). This would suggest a
strongly tensile (positive), rather than compressive, change in the
stress along [001]. In fact, the continuum elasticity  estimates would
yield tensile
stresses along both directions \cite{SanSchEnd98,MeySanPop03}, with the stress along [001] roughly
two times larger.

Moreover, we notice a large change (decrease) in stress along
[1$\bar{1}$0] and almost no change along [001] from the PS  to the  
$1\times7$ configuration, 
although the change in strain (insertion of atoms, i.e., compressive
change in strain) is occurring along [001]. In this case, the
estimated change (decrease) of stress from the elasticity-theory model would be three times
larger along [001], than along [1$\bar{1}$0] \cite{San99}.  Similar
surprising trends in the stress were detected experimentally \cite{MeySanPop03}. 
In fact, there  is a good general agreement between theory and
experiment \cite{SanSchEnd98,MeySanPop03}:  most of the main features
and changes of trends are reproduced, although our calculated change
in the [001] stress from clean W to PS ($-2.0$~N/m) is smaller in
magnitude than the reported measured value in
\cite{SanSchEnd98} ($-2.9$~N/m) and our [001] stress change
from clean W to $1\times7$ ($-1.3$~N/m) is larger/smaller in magnitude
than the measured value reported in Ref.~\cite{SanSchEnd98}
($-0.8$~N/m)/Ref.~\cite{MeySanPop03} ($-2.1$~N/m) \cite{note_temp_ref}.
 It should be noted that
different degrees of coexistence of Ni phases, other than the
reference $1\times7$ phase, are likely to be responsible for the
difference in the experimental values in \cite{SanSchEnd98}
and \cite{MeySanPop03}.

In order to understand the influence of the surface structure on the
stress in Fig.~\ref{fig:stress}, we also  evaluated the surface stress
of the other PS and $1\times7$ structures we considered in the
  previous subsections.  The results
are reported in Table~\ref{table:PS} (for the $1\times1$ structures) and
\ref{table:1x7} (for the $1\times7$ structures). The results for the
$1\times1$  structure in Table~\ref{table:PS} show that the Ni
adsorption site has a critical influence on the surface stress. The H
site (close to T') is characterized by an especially large 
stress  (8.9~N/m) along [1$\bar{1}$0] and low stress (2.9~N/m) along [001], which
account for the stress behaviour of the ground state (T')  PS
structure in Fig~\ref{fig:stress}. The B site, instead, has a larger
stress along [001] (7.5~N/m) than along [1$\bar{1}$0] (5.9~N/m),
whereas the T site is characterized by a more isotropic
(``averaged'') stress ($5.8-6.1$~N/m). The stress is exactly
isotropic somewhere off the T site, towards the Q site, which has
reversed stresses, i.e.,  larger stress (8.4~N/m) is along [001]. At
this site,  along [1${\bar1}$0] the stress (5.1~N/m) is the smallest of all considered $1\times1$
structures, as it is steadily decreasing from the H site to the Q
site.   

Similar trends are
observed for the $1\times7$ structure in Table~\ref{table:1x7}, from
which one can draw a  relation between the stress of the
predominant type of adsorption sites and the stress of the $1\times7$
structures. In particular, the C1 structure is characterized by an
especially large stress (9.5~N/m) along  [1${\bar1}$0] and low stress
(2.7~N/m) along [001]. This is similar to the behaviour of the H
structure, suggesting a dominant influence of the H site in
determining the stress of the C1 structure. Similar to the T
structure, the C2 structure (including many Ni T-like sites) displays a nearly isotropic stress
($4.0-4.1$~N/m). This suggests a dominant influence, in this case, of
T-like sites [Fig.~\ref{fig:1x7_struct} (b)]
on the behaviour of the C2 stress. We also note that the C2 stress is
somewhat lower than the T stress, which is consistent with a higher
density (more compressed) Ni layer in the C2 structure, compared to
the T structure.  For the C3 structure, instead, the
crucial  element for the interpretation of the stress in
Table~\ref{table:1x7} appears to be the  
presence of the 2 Ni atomic rows squeezed into one along [1$\bar{1}$0],
which causes compressive stress along [1$\bar{1}$0].

We can apply the above findings towards an explanation of the anomalous
trend (with respect to
the expectations from the model elasticity theory) observed in
Fig.~\ref{fig:stress} and in the experiment, from PS to
$1\times7$ configurations. From  elasticity theory, 
an essentially constant stress along [1$\bar{1}$0] and a significant decrease of stress along
[001] are expected, while in both experiment and our calculations, we  
observe the opposite: there is a strong decrease along [001] and a
slight increase along [1$\bar{1}$0]. 
The observed trends can be rationalized, however, by considering the
change in Ni adsorption sites, from H-like or T' site
(i.e. essentially two-fold coordinated sites) in the PS configuration
to mainly three-fold coordinated T-like sites in the C2
configuration. In fact, the results in  Table~\ref{table:PS}
demonstrate a major influence of the type of site, occupied by the Ni,
on the stress. They also indicate that the two-fold H-like sites tend
to yield highly anisotropic stresses, with $T_{[1\bar{1}0]}>>
T_{[001]}$, whereas the three-fold T-like sites tend to produce an
isotropic stress, whose 
value is equal to the average over the two perpendicular
directions of the Ni stress components (for the same Ni density).
 On the basis of these considerations, 
the stress from PS to $1\times7$ is expected to
increase along [001] and decrease along  [1$\bar{1}$0], with the
decrease being larger in magnitude, due to the effect of a larger Ni
density in the $1\times7$ than in
the PS case.
We would like to emphasize that, in our case, unlike the case of
alkali-metal adsorbates \cite{MulDahIba02}, the trends in 
 the stress changes are not  driven by a charge 
transfer effect. In fact, we evaluated the work function change (which
is a measure of charge transfer) between  the $1\times1$ and $1\times7$
phases. The work-function
change between the  T' $1\times1$ and C2 configurations is $0.2$~eV. 
The small change of the work function, as calculated, 
indicates a very small charge transfer, which cannot explain the large
change in stress when the Ni site changes \cite{note_wf}.

Meyerheim et al. \cite{MeySanPop03} also  emphasized the
importance of the W's 1$^{st}$-layer displacement for their proposed
structure, which had significant displacements of W atoms. 
We have checked  the influence of the $1^{st}$ W's layer
on the surface stress in the C3 structure, the most disordered
$1\times7$ structure we obtained theoretically. In the proposed structure
based on the SXRD, the W in-plane displacements are extremely large, reaching
even 0.5~\AA. In our structure, the maximum displacements are about 10
times smaller.   We fixed W positions in this layer to the ideal bulk
positions, keeping the interlayer spacing unchanged. The stress of the C3
structure thus changed by only $-0.4$~N/m in both directions, indicating the W displacements
are likely not the cause of
the drastic stress  change of the C3 structure with respect to PS,
and instead it is likely that the influence of Ni atoms
and their configuration is of the essential influence. 

Finally, we note that the  C2 structure  may be defined with two
harmonic functions, which describe the atomic displacements along
[1${\bar1}0$] and [001] \cite{Merwes}, and create nearly regular
hexagons.  The resulting stress
is actually almost isotropic, so one could hypothesize that although
the system is not necessarily relaxing the surface stress, it is
approaching the limit of a clean Ni surface. Even  our calculated
values  are rather close, around 4~N/m for C2 and 3.5~N/m for the
non-magnetic Ni(111).  
Therefore, it seems that in the case of the Ni/W(110) $1\times7$
reconstruction the minimum energy
configuration coincides with highly isotropic arrangement of Ni atoms whose
average stress also corresponds to the clean Ni(111) surface.

\section{Conclusion}

Motivated by puzzling measured stress changes in Ni/W(110), we
performed a first-principles DFT study of the surface stress and
structural properties of the PS and  $1\times7$ coincidence
phase. We determined the ground state $1\times7$
structure which is different than the one proposed based on SXRD. The latter
structure is very irregular, with many distorted Ni hexagons, while
ours includes a highly isotropic Ni(111)-like layer and can be
described by  oscillating Ni[001] chains with their axis projection on
the W surface located mid-way between adjacent rows of [001] W
atoms. The ground state $1\times7$ coincidence structure we
obtain is, instead, similar to the model-theory prediction for the
$1\times8$ Ni/Mo(110) structure and to the experimentally and
theoretically described $1\times8$ phase of Co/W(110). 

Furthermore, our calculated stresses qualitatively follow the measured stress
changes with Ni coverage and reproduce the surprising trend of a larger stress change
along the direction perpendicular to the Ni atomic density change. We
explained the trends in the stress changes in terms of a dominant
influence on stress of the Ni adsorption/coordination sites, as
  oposed to the interpetations based on the continuum elasticity
  theory or charge transfer. We expect
our conclusions concerning the stress dependence on the
adsorption/coordination sites to apply more generally to other related
bimetallic systems involving transition-metal atoms.

\appendix

\section{$1\times7$ atomic positions}

Here we list the atomic coordinates of the upper three layers for the
lowest-energy $1\times7$ configuration, C2.  

\renewcommand{\baselinestretch}{1}
\begin{table}[th]
\caption{Atomic positions in \AA~ for the upper three layers for $1\times
  7$. x is along [1$\bar{1}$0]  and y along   [001].  }
\scriptsize
\begin{center}
\begin{tabular}{ |c|c|c|c|c|c|c|c|c|c|c|c|}
\br
 & $x$ & $y$ & $z$ & & $x$ & $y$ & $z$ & & $x$ & $y$ & $z$ \\
\hline \hline  
 $\rm Ni_1$  & $16.86$ & $10.68$ & $24.12$   & $\rm W_1$  & $19.92$ & $10.99$ & $22.14$  & $\rm W_8$  & $2.20$ & $0.00$ & $19.92$  \\ 
 $\rm Ni_2$  & $ 3.38$ & $0.89$ & $24.20$   & $\rm W_2$  & $2.20$ & $1.57$ & $22.18$  & $\rm W_9$  & $4.42$ & $1.57$ & $19.93$  \\ 
 $\rm Ni_3$  & $5.35$  &  $2.15$ & $24.17$  &  $\rm W_3$  & $4.44$  &  $3.14$ & $22.12$  &  $\rm W_{10}$  & $6.65$  &  $3.14$ & $19.91$  \\
 $\rm Ni_4$  & $ 5.17$ & $4.60$   & $24.06$    & $\rm W_4$  & $6.66$ & $4.70$   & $22.11$  & $\rm W_{11}$  & $8.87$ & $4.71$   & $19.91$  \\
 $\rm Ni_5$  & $7.41$ & $3.35 $   & $24.08$    & $\rm W_5$  & $8.86$ & $6.28 $   & $22.17$  & $\rm W_{12}$  & $11.08$ & $6.28 $   & $19.92$  \\ 
 $\rm Ni_6$  & $11.95$ & $5.81 $   & $24.15$   &  $\rm W_6$  & $11.05$ & $7.84 $   & $22.16$   &  $\rm W_{13}$  & $13.26$ & $7.85 $   & $19.92$  \\ 
 $\rm Ni_7$  & $9.96$ & $7.03 $   & $24.21$   & $\rm W_7$  & $13.26$ & $9.42 $   & $22.12$    & $\rm W_{14}$  & $15.51$ & $9.42 $   & $19.92$    \\
 $\rm Ni_8$  & $12.40$ & $8.25 $   & $24.14$   & $$  & $$ & $ $   & $$  & $$  & $$ & $ $   & $$    \\
 $\rm Ni_9$  & $ 14.72$ & $9.44 $   & $24.08$   &  $$  & $ $ & $ $   & $$    &  $$  & $ $ & $ $   & $$     \\
\br
\end{tabular}
\label{table:1x7_positions}
\end{center}
\end{table}

\ack
We gratefully acknowledge A. Locatelli and J. Ardini  for  
sharing their unpublished data with us and for helpful discussions. We are very
thankful to E. Bauer for his critical reading of the manuscript. We also
acknowledge T.~O. Mente\c{s} for fruitful discussions. The
calculations were performed on the sp6 supercomputer at CINECA. 

\section*{References}
\bibliography{NIW}

\begin{thebibliography}{10}
\bibitem{note_W}
We note that our calculated W(110) relaxations are given in \cite{MenStoBin08}
  and that the first two relaxations are identical for the 5-layer W film used
  in the present work and a thicker film from \cite{MenStoBin08}.
\bibitem{note_shear}
Almost all Ni/W(110) structures considered in this work (the only exception is
  the H-site PS structure) have also some shear stress components, the
  magnitudes of which are always smaller than 1~N/m and of which the largest
  one is for the C3 structure.
\bibitem{note_temp_ref}
We note that our calculations give results for $T= 0$~K, and one could expect
  some temperature effect on stress (the measurements in
  \cite{SanSchEnd98,MeySanPop03} were done at room temperature)
  \cite{StoMenBin10}. However, we expect them to be small, as we have evidence
  from LEED I-V that our and the experimental room-temperature PS structure are
  very similar \cite{note_Jacopo}.
\bibitem{note_wf}
We also evaluated the work function change between the clean W(110) and
  $1\times1$ phases: the change is $-0.4$~eV for the T' and H sites and
  $-0.6$~eV for the B site. This is almost identical to the previous LDA
  \cite{LeuKaoSu03} and GGA \cite{HuaChaLeu05} findings and somewhat different
  (by 0.3-0.4~eV in magnitude) from the experimental value \cite{KolBau85}.
  However, from both theoretical and experimental value of the work-function
  change, which is considerably smaller in magnitude than for Li/Mo(110)
  \cite{MulDahIba02}, one can deduce that the charge transfer is not
  responsible for the drastic surface-stress changes.
\end{thebibliography}

\end{document}